\documentclass[preprint,aps,tightenlines,showpacs]{revtex4}
\usepackage{amsmath,amsfonts,amssymb,amsthm} 
\usepackage{color}
\usepackage{graphicx,latexsym}

\begin{document}
 \title{ Theory of the Energy Variance in  a Quantum Bit}
\author{Gilbert  Reinisch}
\email{Gilbert.Reinisch@oca.eu}
\affiliation{Universit\'e de la C\^ote d'Azur - Observatoire de la C\^ote d'Azur\\
06304 Nice Cedex - France}
\author{and}
\email{gilbert@hi.is}
\affiliation{Science Institute, University of Iceland, Dunhaga 3, IS-107 Reykjavik,
             Iceland}

\begin{abstract}
The energy  of a  driven  two-level quantum system ---or  quantum bit (qubit)--- can be 
exactly  (i.e. without the use of the rotating wave approximation)  defined by  the time derivatives of
its 
large-amplitude,    high-frequency  internal
as well as   overall phases.  We show that their envelope can be  derived  from 
the quantum   expectation  value $\langle\mathbf {V} \rangle=\langle \Psi |\mathbf {V} |\Psi \rangle$  of 
 the   time-dependent Hermitian  variance operator
$ \mathbf {V}=[\mathbf {H} -\langle\mathbf {H} \rangle\mathbf {I}]^2$.  Indeed, following standard statistical physics, 
we formally define   this
 latter  by  the use of
 the Hamiltonian $\mathbf {H} $,  the state $\Psi $       and the    energy mean  value 
$\langle\mathbf {H} \rangle =\langle{\Psi} |\mathbf {H} |{\Psi} \rangle$ of the system 
($\mathbf {I}$ being  the identity matrix).
Remarkably,  the   resulting
     standard deviation     $\sigma =\sqrt{\langle\mathbf {V} \rangle}$
may become comparable to   the energy mean  value itself. We have indeed $\sigma \sim \langle\mathbf {H} \rangle$
in the case of a weak Rabi drive. Therefore  $\sigma $
yields    a large statistical time-dependent     range  of available  values
for the qubit energy about its  well-known harmonic Rabi state flipping   $\langle\mathbf {H} \rangle$.
By which experimental protocol  can such a intriguing quantum property be  verified?  
In
   the spirit of a recent comment (\emph{I. Mazin, Nature Physics 2022,  18, 367}),
we point out  the
experimental measurement of the so-called
 ``time-of-flight''    values of an abrupt  quantum jump  (\emph{Z. Minev et al. Nature 2019, 570, 200})
as a possible candidate for  such a protocol. We show that we quite simply 
 recover  these  abrupt  quantum jumps  by use of both the present theory and the quantum Zeno effect.
\end{abstract}


   \maketitle

 \subsection{Introduction}

How can the energy variance of
 a two-state quantum system  ---conservative or not--- be defined and what is the physical meaning
of its
 corresponding standard deviation? To answer these questions, the present paper proposes  a rigourous  ---no rotating wave approximation (RWA)---
  quantum theory  that formally duplicates the  
 standard   definition of 
  variance in statistical physics. It   shows through the discussion of
a recent experiment \cite{Minev19} how such a theoretical background can actually 
apply. 

Adopting the bra-ket formalism, one   first
recalls
 the standard quantum expectation --or ``average'', or ``mean''--
value:
\begin{equation}
\label{Hmean}
\langle \mathbf {H}   \rangle  = \langle{\Psi} |\mathbf {H} |{\Psi} \rangle 
\end{equation}
of the energy   by use of  the Hamitonian $\mathbf {H} $ of the system
and its  wavefunction $\Psi $. Then, one might wish to introduce the following
variance  operator ($\mathbf {I}$ is the identity matrix):
\begin{equation}
\label{QuantumVarianceOP}
  \mathbf {V} =[\mathbf {H} -\langle \mathbf {H}   \rangle  \mathbf {I}]^2 ,
\end{equation}
that is directly extrapolated from standard statistical physics. This operator is Hermitian with two  real eigenvalues. 
 The resulting variance expectation value:
\begin{equation}
\label{QuantumVariance}
  \langle\mathbf {V}  \rangle= \langle{\Psi} |  \mathbf {V}  |{\Psi} \rangle,
\end{equation}
yields the corresponding standard deviation:
\begin{equation}
\label{QuantumStandardDeviation}
\mathbf {\sigma}  =\sqrt{  \langle\mathbf {V}  \rangle} 
\end{equation}
of the energy with respect to  
its mean-value    (\ref{Hmean}). 
How can such a formal dispersion of energy values about  $\langle \mathbf {H}   \rangle $
 be measured?

In order to  answer this basic question, we refer to the  exact solution of the Hamiltonian equations of motion
 of a driven qubit   \cite{reinisch1998}    \cite{Reinisch98}. Since the RWA is  discarded, one keeps the
two high-frequency energy components $E_{a,b}(t)$ related to  the qubit's energy mean value (\ref{Hmean}) by:
\begin{equation}
\label{EnMoyab}
\langle \mathbf {H(t)}   \rangle=|\psi_a(t)|^2 E_a(t)+|\psi_b(t)|^2 E_b(t)  .
\end{equation}
Here  the so-called ``state energies'' $E_{a,b}(t)$  are defined in the next section by  the time derivatives of
the qubit's     internal
and    overall phases while
$\psi_{a,b}(t)$ are the   spinor components of its wavefunction $\Psi(t)$. We assume   that its  Hamiltonian 
$\mathbf {H(t)}$
is defined by the external time-dependent energy drive ${\cal E}(t)$ and by the  off-diagonal constant $K$
through the two  Pauli matrices as:
\begin{equation}
\label{hamiltTWOstates}
{\Psi}(t) =
\begin{pmatrix}
   \psi_a(t) \\
 \psi_b(t)
\end{pmatrix}\quad;\quad
\mathbf {H}(t)
=
\begin{pmatrix}
    {\cal E}(t) &{K} \\
  { K} & -  {\cal E}(t) 
\end{pmatrix} .
\end{equation}
We show  that the envelope of   the state energies   $E_{a,b}(t)$   
is quite accurately determined  by the quantum standard deviation (\ref{QuantumStandardDeviation}).
 Since they
 are simply phase-shifted by $\pi$ \cite{reinisch1998}, they are almost redundant. Therefore their mean value within  their   high-frequency period $T$ is  equal to 
 $\langle \mathbf {H}   \rangle $ as shown by eq.  (\ref{EnMoyab}) \cite{reinisch1998}.
This   explains the role played by $\mathbf {\sigma}$ in an energy measurement process. Indeed, if the  measurement  at time $t$   lasts less than
$T$, it yields  either of the two state energies  $E_{a,b}(t)$. Conversely, if it lasts more than $T$, it 
yields    $\langle \mathbf {H}\rangle $ and no special effect due to $\mathbf {\sigma}$ can be observed. 
Therefore the  standard deviation (\ref{QuantumStandardDeviation}) will play a leading role in any series of  energy measurements
whose frequency is higher than   $1/T$. The quantum Zeno effect (QZE)  hence appears as  the ideal  candidate  for  an experimental check of the present theory because  it     demands    frequent (almost continuous) measurements \cite{Itano19}.
Then QZE
  inhibits  the decay of  an
unstable quantum system through the
wave-function collapse described by the quantum measurement theory
 \cite{Neumann32}:  the system      evolves
from the same initial state after every measurement \cite{Misra77}\cite{Home93}.

This effect has been beautifully 
verified  by use of  a Rabi-driven two-level quantum system \cite{Itano90} ---namely, a ground state |G>
 and an excited state |D>---  with the addition of  a 3rd ``ancilla'' state |B> that actually
plays the role of the  continuously-operating ground-state population measurement  \cite{Cook88}. 
Specifically,   state |B> is connected by a strongly allowed transition
to level |G> and it
can  decay only to  |G>. The continuous  state measurement is carried out
by
 resonantly (Rabi) driving the $ G \rightarrow B$ transition with an appropriately designed optical pulse.
This measurement  causes a collapse of the wave function as told above.
If the system 
is  projected into the ground-state level |G> at the beginning of the pulse, it cycles between
|G> and |B> and emits a series of photons ---hence the
label |B>  for  ``bright''--- until the pulse is turned off. If it is projected into the excited level |D> (for ``dark''), it scatters no photons.
Therefore the wave-function collapse is due to a null measurement \cite{Porrati87} \cite{Raymond2020}. That is, 
the \emph{absence} of scattered photons when the optical pulse is applied is enough to cause a
collapse of the wave function to level |D>  \cite{Itano90}.

A recent outstanding  experiment has reproduced this 
experimental set-up \cite{Minev19}.   It  led to  the occurence of  sudden, quite abrupt  quantum jumps.
 The agreement between
the observation and the theoretical predictions ---based on the   arduous
 modern quantum trajectory theory \cite{MinevPHD18}--- is 
excellent. In order to put to the test the present  theory,
we wish to recover the existence of these abrupt  quantum jumps by a
quite  simple alternative explanation  derived  from   the above quantum-variance properties.
We believe that this      scope  is in agreement with
 the spirit stressed  in   \cite{Mazin2022}.
 \subsection{Theory}
 In the normalized  two-state $|a,b\rangle$  base,  the Schr{\"o}dinger  equation:
\begin{equation}
\label{schroeTWOstates}
i \hbar\frac{\mathrm{d}}{\mathrm{d} t}
{\Psi} 
= 
 \mathbf {H}
{\Psi} 
\end{equation}
defined by Eqs (\ref{hamiltTWOstates})
is equivalent to
the following Hamiltonian formulation of classical-like mechanics
 by use of the two conjugate  canonical coordinates  $\delta(t)$ and $\alpha(t)$:
\begin{equation}
\label{hds}
\dot{\alpha}=-\frac{\partial {\cal H}}{\partial \delta} = \sqrt{1-\alpha^2} \sin \delta
\quad ; \quad
\dot{\delta}=\frac{\partial {\cal H}}{\partial \alpha} = -\frac{\alpha}{\sqrt{1-\alpha^2} } \cos \delta + {\cal E}(\tau) ,
\end{equation}
  whose         Hamiltonian 
is:
\begin{equation}
\label{Hhds}
{\cal H} = \sqrt{1-\alpha^2} \cos \delta + \alpha {\cal E}(\tau) , 
\end{equation}
provided \cite{reinisch1998}    \cite{Reinisch98}:
\begin{equation}
\label{fctdondesTWOstates}
\psi_a(t) =\sqrt{\frac{1+\alpha(t)}{2}} e^{i \Theta(t)}  \quad ; \quad
\psi_b(t) =\sqrt{\frac{1-\alpha(t)}{2}} e^{ i[ \Theta(t) +\delta(t)]} .
\end{equation}
In the above  (\ref{hds}-\ref{Hhds})  Hamiltonian dynamical system (or HDS), the dot means the derivation with respect to
the dimensionless time $\tau=\Omega t$  with
 $\Omega = 2{K}/\hbar$ (e.g. it is the Larmor
frequency  of a two-level spin one-half system).
Hence  we assume in the present work:
\begin{equation}
\label{paramNonDim}
{K}=1\quad ; \quad \hbar=2\quad ; \quad \Omega =1 .
\end{equation}
The angular positions  $\theta$  and $\phi$  on the qubit's  Bloch sphere are
straightforward:
\begin{equation}
\label{Bloch1}
\alpha=\cos\theta\quad;\quad\phi=\Theta + \frac{\delta}{2}.
\end{equation}
Then:
\begin{equation}
\label{Bloch2}
\psi_a(t) =e^{i\phi(t)} \cos\frac{\theta(t)}{2}  e^{-i\delta(t)/2}  \quad;\quad 
\psi_b(t) =e^{i\phi(t)} \sin\frac{\theta(t)}{2}  e^{+i\delta(t)/2} .
\end{equation}
Note that the time-dependent
 overall phase $ \Theta(\tau)$  in (\ref{fctdondesTWOstates})
is  \emph{not} a 3rd independent variable: it is slaved to the solution of  HDS 
(\ref{hds}-\ref{Hhds})
by:
\begin{equation}
\label{ThetaDOT}
\dot{\Theta}=-\frac{1}{2}\Bigl[\sqrt{\frac{1-\alpha}{1+\alpha}}\cos\delta+{\cal E}(\tau)\Bigr] .
\end{equation}
Contrary
to intuitive  opinions, the dynamics of the overall
phase of a quantum state can indeed yield an observable
physical effect (e.g. it  causes the famous $4\pi$-symmetry
of spinor wave functions that have been directly
verified in both division-of-amplitude \cite{Rauch75}\cite{Werner75} and
division-of-wave-front \cite{Klein76} neutron interferometry
experiments).
When $\alpha\equiv 0$ 
(which is Feynman's  rough  assumption  in  
his textbook  description of the stationary Josephson effect \cite{Feynman1965} ), 
Eqs. (\ref{hds})  yield the two Josephson equations 
(${\cal E}$ is then the applied voltage).

In the simplest 
conservative case ${\cal E}\equiv 0$, 
the classical-like HDS orbits defined by   Hamilton equations (\ref{hds}) do all have    the same reduced frequency +1 (actually $+\Omega$) for ${\cal H}>0$ (resp. -1, or $-\Omega$, for ${\cal H} < 0$). They
  define 
the corresponding quantum  superposition states of the system
in agreement with Eqs. (\ref{fctdondesTWOstates}) and  (\ref{ThetaDOT}) \cite{reinisch1998}.
 Such a binary structure 
of the orbit frequency, namely $\pm\Omega$ in actual units, is the translation 
in terms of the HDS equations of motion (\ref{hds}-\ref{Hhds})  and  action \cite{reinisch2021a}, of the eigenvalues $\pm 1$ in  the two-level energy 
spectrum of the undriven  qubit.
In  the driven, however  still conservative case   ${\cal E} \equiv constant \ne  0$,
all the orbits  do still  keep the same angular frequency $\pm\sqrt{1+{\cal E}^2}$ (i.e.  the corresponding
eigenvalues),
depending on the sign of ${\cal H}$. Moreover it has also  been shown in  \cite{reinisch1998} that
 the above-mentioned $4\pi$ symmetry of the overall qubit phase $\Theta$ 
becomes  an immediate consequence of the HDS solution of Eq. \ (\ref{ThetaDOT}).

We have in accordance with Eqs.\ (\ref{Hmean}),
 (\ref{hamiltTWOstates}),  (\ref{Hhds}) and  (\ref{fctdondesTWOstates})    
the following fundamental  link between the mean qubit energy $\langle \mathbf {H}   \rangle$ and the HDS energy  ${\cal H}$ in the dimensionless units (\ref{paramNonDim}):
\begin{equation}
\label{Hmeanbis}
\langle \mathbf {H}(\tau)    \rangle= {\cal H}(\tau) .
\end{equation}
Equations  (\ref{EnMoyab}-\ref{fctdondesTWOstates}) and  (\ref{Hmeanbis})
yield
the following HDS definition of the  two qubit's  time-dependent  \emph{state energies} in the 
dimensionless units (\ref{paramNonDim}):
\begin{equation}
\label{Epsab}
 E_a   =  \frac{{\cal H}+{\cal E}}{1+\alpha} \quad ; \quad E_b   =  \frac{{\cal H}-{\cal E}}{1-\alpha} .
\end{equation}
They are the derivatives of  the two internal as well as overall  phases  $\delta(\tau)$ and $\Theta(\tau)$:
\begin{equation}
\label{EpsabBIS}
  E_a
   = - \hbar 
\frac{\mathrm{d}\Theta}{\mathrm{d} t} = -2  {\dot \Theta} \quad ; \quad 
 E_b = - \hbar 
\frac{\mathrm{d}(\Theta+\delta)}{\mathrm{d} t}  =   -2  (\dot {\Theta}+ \dot {\delta})   ,
\end{equation}
by use of  Eqs. (\ref{hds}) and (\ref{ThetaDOT}). They provide 
an explicit  fine-structure  mapping of 
  the dynamical properties of the classical-like  HDS trajectories 
 (\ref{hds}-\ref{Hhds})  onto the mean quantum  energy (\ref{EnMoyab})  
and   onto its  state energy  variance:
\begin{equation}
\label{sigmaab}
 \sigma_{a,b}^2(\tau)  =   |\psi_{a,b}(\tau)|^2 [E_{a,b}(\tau)-{\cal H}(\tau)]^2  ,  
\end{equation}
in terms of  the
    standard deviations $\pm \sigma_{a,b}(\tau) $.
The  necessary  link  with the quantum variance (\ref{QuantumVarianceOP}-\ref{QuantumVariance})
is  performed as follows. Using (\ref{Hmeanbis}), we have:
\begin{equation}
\label{varianceOP}
 \mathbf {V}   =   
\begin{pmatrix}
    ({\cal H}-{\cal E})^2+K^2 &-2K{H} \\
 -2K{H} &({\cal H}+{\cal E})^2+K^2 
\end{pmatrix} ,  
\end{equation}
and therefore, by use of Eqs. (\ref{QuantumVariance}) and (\ref{fctdondesTWOstates}):
\begin{equation}
\label{meanVariance}
 \langle \mathbf {V}   \rangle  =   (1-2K) {\cal H}^2 + 2 (K-1)\alpha {\cal E} {\cal H} + {\cal E}^2+  K^2.  
\end{equation}
Since we assume $K=1$ in accordance with (\ref{paramNonDim}), Eq. (\ref{meanVariance})
becomes:
\begin{equation}
\label{meanVarianceAPPROX}
 \langle \mathbf {V}   \rangle  =   {\cal E}^2 - {\cal H}^2 +1 \sim 1-{\cal H}^2,
\end{equation}
 if the qubit is weakly Rabi-driven as envisaged  in the next section.  Then the link between the quantum-variance 
standard deviation (\ref{QuantumStandardDeviation}) and  the state-energy ones (\ref{sigmaab})
becomes clear: see Fig. \ref{HrdPLUSsig}.
 \subsection{Weak  Rabi drive}
When the   weak resonant field: 
\begin{equation}
\label{WeakDrive}
 {\cal E}(\tau)=A\sin\tau \quad ; \quad A\ll1,
\end{equation}
  is applied to the system, we obtain
 in accordance with Eq. (\ref{Hhds})
 the
well-known  quasi-harmonic  Rabi  oscillation between the   two eigenvalues $\pm 1$ of the mean  energy:
\begin{equation}
\label{meanHrdANAT}
 {\cal H} \sim \cos\frac{A\tau}{2} +\Delta  {\cal H}  \quad ; \quad \Delta  {\cal H} =  \alpha {\cal E}(\tau) =  A\alpha \sin\tau \ll 1 .
\end{equation}
\begin{figure} 
 \includegraphics[width=0.65\textwidth,angle=0,clip]{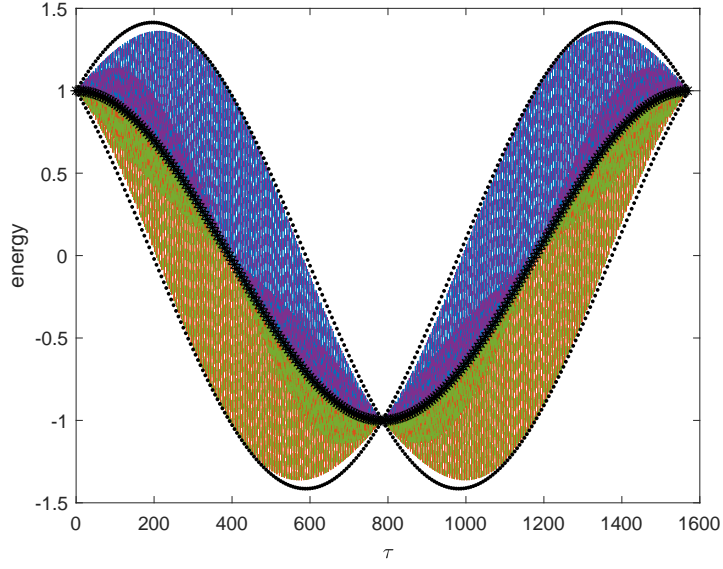}
      \caption{Validation of the state-energy description
(\ref{Epsab}-\ref{sigmaab}) by use of  the variance Hermitian operator  (\ref{QuantumVarianceOP}-\ref{QuantumVariance}) 
for
$A=8\,10^{-3}$. The
qubit's colored   energy  regions   are  defined by the state-energy standard deviation
dressing (\ref{sigmaab}) of  Rabi's quasi-harmonic HDS Hamiltonian ---or mean energy---  (\ref{meanHrdANAT}) 
 during  one  Rabi period $4\pi/A=1571$ in reduced units (\ref{paramNonDim}) (thick black plot). 
Dotted black plot: the quantum standard deviations (\ref{QuantumStandardDeviation}) about ${\cal H}$
as defined by quantum variance (\ref{meanVarianceAPPROX}).}
\label{HrdPLUSsig}
\end{figure}
One obtains a regime of HDS quasiperiodic orbits slowly spiraling out of one ${\cal H}=\pm1$ eigenvalue  cell  og in to the next 
${\cal H}=\mp 1$ one \cite{Reinisch98}.
 The corresponding ``dressing'' of the quasi-harmonic  mean energy  value (\ref{meanHrdANAT}) 
by the state-energy standard deviations defined by   variance (\ref{sigmaab}) is quite spectacular  \cite{reinisch2021a}. In Fig. \ref {HrdPLUSsig},
 ${\cal H}(\tau)$ is displayed
 by  the  thick black plot while   the quantum standard deviations about it,
 defined by Eq. (\ref{QuantumStandardDeviation})  and  by quantum variance (\ref{meanVariance}), are shown by the dotted black plots.
The quite dense   colored patterns  enclosed by these latter
are built from
the very-high-frequency oscillatory standard deviations  (\ref{sigmaab}) with  extremely small  HDS orbit  period $\sim 2 \pi \ll  1571$
at the scale of  the Rabi period 
 $4\pi/A=1571$. The   standard deviations 
 $+\sigma_a(\tau)$ (continuous purple) and $-\sigma_a(\tau)$ (continuous green)  due to state energy $E_a(\tau)$ defined  in (\ref{Epsab})
are phase-shifted by $\sim \pi$ 
with respect to the  standard deviations $+\sigma_b(\tau)$ (dotted purple) and $-\sigma_b(\tau)$ (dotted green)
 due to state energy $E_b(\tau)$.
All these  very-high-frequency standard deviations display the rather important dispersion of the qubit   energy  
about its mean value ${\cal H}$  except at the two eigenvalues $\pm 1$ where, as expected, this dispersion vanishes.
We note with interest  that these  standard deviation patterns   are 
quite accurately  bounded
by
the quantum standard deviations 
 (\ref{QuantumStandardDeviation}) defined by the quantum variance (\ref{meanVariance}).
This remarkable property establishes the   statistical
 link between  the variance Hermitian operator description    (\ref{QuantumVarianceOP}-\ref{QuantumVariance}) and
the state-energy one
(\ref{Epsab}-\ref{sigmaab}). It  validates  the self-consistency of the present theory.

\begin{figure} 
 \includegraphics[width=0.65\textwidth,angle=0,clip]{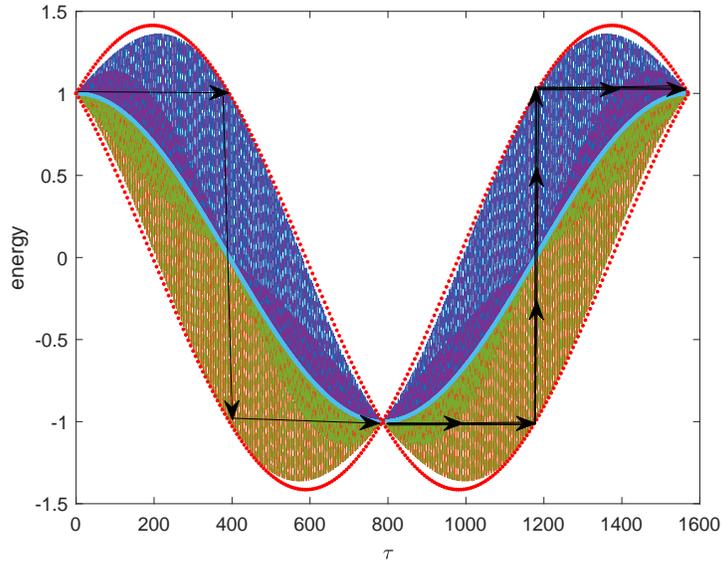}
      \caption{$A=8\,10^{-3}$: the  qubit   energy  regions  defined by Fig. \ref{HrdPLUSsig}
together with the   quasi-harmonic HDS Hamiltonian ${\cal H}$ (in continuous blue)   and  the
quantum standard deviations  defined by Eqs \  (\ref{QuantumStandardDeviation})  
and   (\ref{meanVariance})   about ${\cal H}$ (red dotted plots). The
 general   principle of a  step-like   quantum jump due to  the series of
repetitive  eigenstate measurements   is illustrated by arrows.
 The thick arrows display the  abrupt     energy jump  occuring at $\tau = 1183$
(actually at  ${\cal H}=0$)
from the ground state $E=-1$
to the excited level  $E=+1$.
The thiner path displays the reverse de-excitation  process.}
\label{HrdPLUSsig1}
\end{figure}
 \subsection{Quantum Zeno jump}
The measured ``'time-of-flight'' values   obtained by  a   continuous measurement process
 in experiment    
\cite{Minev19}
have been recovered in the frame of the above theory by  considering the quantization of the  HDS action \cite{reinisch2021a}.
Here we wish to show in addition  that these values are actually an immediate    first-principle consequence 
of
Fig.  \ref{HrdPLUSsig} and ---contrary to the quantum trajectory description  given in \cite{Minev19}--- they
do not depend on the specific continuous measurement process (provided this latter is long-lasting enough). Indeed 
  \emph{any} such   process yields   QZE ``freezing'' of the
corresponding quantum states in their  initial configuration  \cite{Itano19}  \cite{Misra77}
and this freezing  remains  possible as long as the resulting  trajectory lies inside 
of the  qubit's    energy  region  defined by Fig. \ref{HrdPLUSsig}.

Let us be specific and consider Fig. 
\ref{HrdPLUSsig1}.  Assume that the system,
while still resonantly Rabi-driven by (\ref{WeakDrive}), is initially in, say, its ground state $-1$ (i.e. at half the Rabi
period $\tau = 785$) and introduce a repetitive measurement process in accordance with
\cite{Itano90} or, equivalently, with \cite{Minev19} (the following applies as well when  we start from  the excited state: see Fig. 
\ref{HrdPLUSsig1} at $\tau = 0$).
Then, due to QZE which acts as a \emph{strong perturbation}, the system is forced to remain “frozen” in this
lowest energy value instead of starting  Rabi’s  harmonic excitation
dynamics (\ref{meanHrdANAT})  (displayed by the blue line). Consequently, its energy  keeps its value $-1$  along the horizontal segment
   as $\tau$  increases for $785 < \tau < 1183$ (lowest horizontal arrow path), i.e. as long as
the energy of the system  stays within the statistically accessible  energy region about ${\cal H}$
that is defined by Eq.   (\ref{sigmaab}), or equivalently by 
Eqs\ (\ref{QuantumStandardDeviation})  and   (\ref{meanVariance}) (dotted red line). When reaching this  boundary
at $\tau = 1183$, it  jumps to the excited level $+1$ in accordance with
the vertical arrow path in order to continue  the QZE state freezing process from $\tau=1183$ to $\tau=1571$
if the repetitive  measurement is lasting over a time interval greater than this half Rabi period.
This  QZE jump ---which we call ``\emph{quantum Zeno jump}'' (QZJ)--- actually absorbs  at once all the energy input due to the 
external Rabi drive that has been  stored   in 
the system  during its  forced  QZE freezing stage in its ground state.
\begin{figure} 
 \includegraphics[width=0.65\textwidth,clip]{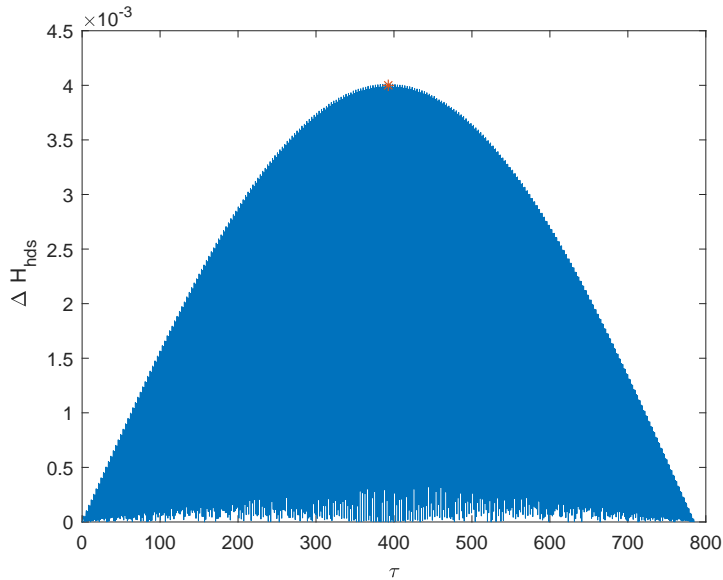}
      \caption{The quasi-stochastic   energy input  $ \Delta  {\cal H}$ defined by Eq. (\ref{meanHrdANAT}) 
over half a Rabi period $2\pi/A$  when the origin of time is now
 taken at $\tau=785$  in
  Figs.  \ref{HrdPLUSsig}-\ref{HrdPLUSsig1}. 
Its seemingly      two-dimensional dense   pattern
results from its extremely fast oscillations  at the scale of the Rabi period. Therefore it
will be  regarded as the  uncertainty in the energy input at a given time $\tau$.
The star indicates the 
maximum  at   $\tau =393$ or
$ {\cal H}=0$ that scales this  uncertainty in  (\ref{PPEheisenberg}).} 
\label{Heisenberg1}
\end{figure}
\begin{figure} 
 \includegraphics[width=0.65\textwidth,angle=0,clip]{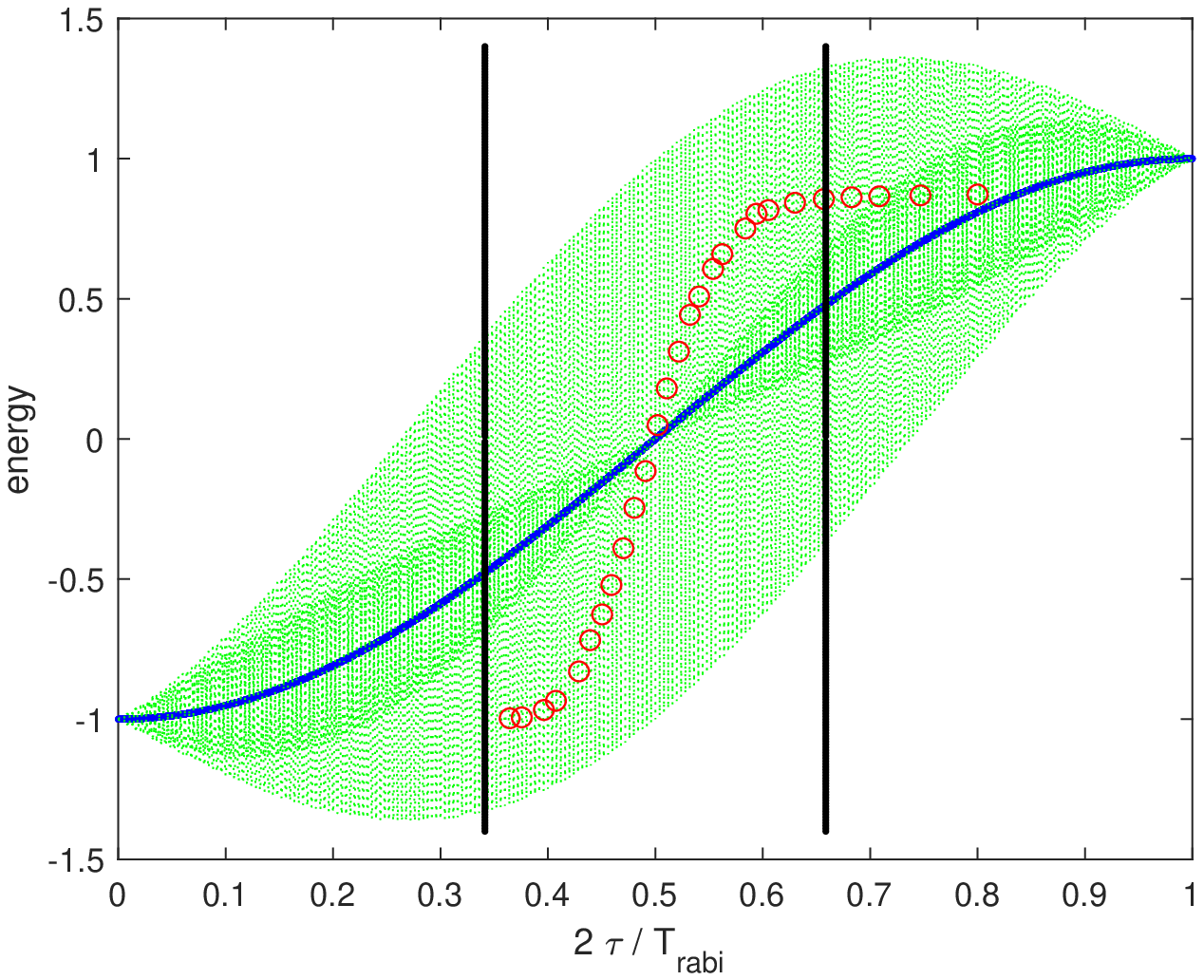}
      \caption{Second (r.h.s.) part of Figs.  \ref{HrdPLUSsig}-\ref{HrdPLUSsig1} illustrating
 the 
\emph{QZE-vs-Rabi}  excitation process of Fig. \ref{HrdPLUSsig1}
by use of the data provided with  \cite{Minev19}.
  Blue plot: Rabi's mean energy defined  by Eq. (\ref{meanHrdANAT}).
Circles:  the
 path in the
Hilbert space followed by the jump evolutions given in  Fig. 3b of  Ref. \cite{Minev19},
using its  $T_{Rabi}=50\,\mu$s   Rabi period  for the normalization of the $\tau$-axis.
The $13\%$ deviation from   the excited eigenvalue +1 is  due  to
imperfections, mostly excitations to higher levels  \cite{Minev19}.
The two vertical bars display  the minimum duration $ \Delta \tau_{min}$ of the transition obtained as the  consequence of 
the uncertainty  principle (\ref{PPEheisenberg})
when the perturbation $\Delta {\cal H}$ illustrated by Fig. \ref{Heisenberg1}   is maximum (star).
These bars  fit quite well with the duration of the so-called “time-of-flight” value $\sim 2\Delta_{mid}$
  given in   \cite{Minev19}. Therefore they experimentally validate the present generic theory.}
\label{Heisenberg3}
\end{figure}

The above QZJ  scheme is  oversimplified. Indeed the standard deviation boundary at  $\tau=1183$ is only statistical: it has not a precise definite value. Therefore the QZJ may occur at any time
in the    interval  
$\Delta \tau $ \emph{about} $\tau=1183$.
When trying to  evaluate it,  one should 
regard the Hamiltonian driving term $\Delta {\cal H} $ in (\ref{meanHrdANAT})
as a quasi-stochastic   energy input,
 as shown by Fig. \ref{Heisenberg1}.
  Indeed, $\Delta {\cal H} $  oscillates extremely rapidly at
the scale of the Rabi period $4\pi/A$. It actually looks like a mere two-dimensional  uniformly-dense
drive   pattern and thus yields
   the corresponding time interval $\Delta \tau$:
\begin{equation}
\label{PPEheisenberg}
 \Delta {\cal H}  \Delta \tau \geq \frac{\hbar}{2}=1 ,
\end{equation}
(cf. our choice (\ref{paramNonDim})
of the  reduced units)  in which any step-like QZJ illustrated by Fig. \ref{HrdPLUSsig1}
can
statistically occur. Since the   QZJ formely appears at half the
gap defined by  $  {\cal H}=0$  (cf. Fig. \ref{HrdPLUSsig1}), we take 
$ \Delta {\cal H}= \Delta {\cal H}_{max}=4\,10^{-3}$ in (\ref{PPEheisenberg})
---see the star
in Fig. \ref{Heisenberg1}— and therefore 
$\Delta \tau_{min}\sim 1/\Delta {\cal H}_{max} \sim 250$.
This time interval 
is pictured by the two vertical bars  
in Fig. \ref{Heisenberg3} where the experimental data given in  Fig. 3b of  Ref. \cite{Minev19}
 are reproduced by use of   red circles, using for the normalization of our $\tau$-axis the experimental Rabi period
$T_{Rabi}= 50\,\mu s$
given in
\cite{Minev19}  (their $\sim 13\%$ deviation from the excited eigenvalue +1 is due, the authors say, to
imperfections, mostly excitations to higher levels). 
We see that $\Delta \tau_{min}\sim  250\sim \sim 2\Delta_{mid}$   agrees fairly well with
the so-called “time-of-flight” value $\sim 2\Delta_{mid}$
  given   in  Fig. 3b of  \cite{Minev19}. Moreover, $\Delta \tau_{min}$ is also in good
agreement with   HDS action quantization when the system crosses the separatrix of the
system at  $ {\cal H}=0$  \cite{reinisch2021a}. Recall   that  Heisenberg's  l.h.s. of inequality (\ref{PPEheisenberg}) has indeed the dimension of an
action.
 \subsection{Conclusion}
The present work     defines  a new quantum Hermitian operator ---the energy variance operator $\mathbf {V} $--- which is simply duplicated
 from the   statistical definition
of   energy  variance in classical physics.  Its  expectation value yields  the standard deviation $\sigma$ of the energy  
about its mean value $\langle \mathbf {H}   \rangle$. We show by use of an exact Hamiltonian description    that $\sigma$  is actually due to the very-high-frequeny energy oscillations
about $\langle \mathbf {H}   \rangle$ which are usually discarded  in the rotating wave aproximation (RWA). Therefore the present 
theory 
 restores
the fundamental physical interest of these latter: while the RWA yields  quite  accurately the Rabi oscillations of 
$\langle \mathbf {H}   \rangle$, the    standard deviation $\sigma$  about $\langle \mathbf {H}   \rangle$
is a mere signature  of the   existence of very-high-frequeny energy oscillations $E_{a,b}$   in  the system.  As a consequence,
the experimental effect of $\sigma$ will be important only within their period while it will average to $\langle \mathbf {H}   \rangle$ 
when the measurement process lasts more than such a period. In the present state of the art, the quantum Zeno effect (QZE) seems to be
the ideal experimental candidate for the detection of $\sigma$  since  it     is due to  a series of   high-frequency measurements whose period may well
become  less than that of the energy oscillations  $E_{a,b}$. This  is indeed the case in  the experimental
measurements of the duration of   quantum jumps   \cite{Minev19} since  they  were actually
performed by reproduction of the original set-up 
that led  to  the experimental discovery of QZE  \cite{Itano90}. The agreement of our results
with those
displayed in \cite{Minev19} constitutes a clear evidence of the physical interest of the standard deviation,
and hence of the variance operator $\mathbf {V} $, in the definition  of the energy of the system.

 \noindent Two final comments:\hfill\break
- We considered a weak resonant Rabi drive  in order to   compare
with the  experimental results found in  \cite{Minev19}. The case of  a strong    Rabi drive has been considered in 
\cite{reinisch2022a}. It was shown there that  the definition of the state energies energies $E_{a,b}$
still remains valid. Their spectrum however may appear either chaotic or   discrete (frequency combs) due to 
phase-locked cycles in the solution of the Hamiltonian  differential system. \hfill\break
- The present definition of the energy variance operator $\mathbf {V} $ may actually be generalized to any observable operator
$\mathbf {O}$.
Then its variance operator will simply be  $ \mathbf {V[O]= [O - \langle O\rangle I]^2}$  and its resulting expectation value
$\langle  \mathbf {V[O]} \rangle$
in a given state $\Psi$
will define the corresponding standard deviation about  $\langle  \mathbf {O}\rangle$  in any measurement process that takes into account the high-frequency
components of the Schr{\"o}dinger solution.

\bibliography{mod_qd}
%

\begin{acknowledgments}
 \noindent   The author wishes to thank D. Esteve for his sharp and useful comments about the physics concerned by   the present work.
He  
 gratefully
acknowledges financial and technical  support from UMR \emph{Lagrange} (Observatoire de la C\^ote d'Azur, universit\'e de Nice, France) as well as  from the Icelandic
      Technology Development Fund (University of Iceland, Reykjavik, Iceland).
\end{acknowledgments}


\end{document}